\documentclass[superscriptaddress,prl,twocolumn,nofootinbib,showpacs,preprintnumbers]{revtex4}

\usepackage{graphicx,epsfig}

\def\bwt{\begin{widetext}}

\def\ewt{\end{widetext}}

\def\be{\begin{equation}}

\def\ee{\end{equation}}

\def\bea{\begin{eqnarray}}

\def\eea{\end{eqnarray}}

\def\bean{\begin{eqnarray*}}

\def\eean{\end{eqnarray*}}

\def\bary{\begin{array}}

\def\eary{\end{array}}

\def\bit{\begin{itemize}}

\def\eit{\end{itemize}}

\def\su5u1{SU(5) \times U(1)}

\def\fsu5u1{SU(5) \times U(1)'}

\def\so10{SO(10)}

\def\sq20{SO(10) \times SO(10)}

\usepackage[centertags]{amsmath}

\usepackage{amssymb}

\newcommand{\Z}{{\mathbb Z}}

\begin{document}

\title{A Realistic World from Intersecting D6-Branes}

\author{Ching-Ming Chen}

\affiliation{George P. and Cynthia W. Mitchell Institute for
Fundamental Physics, Texas A$\&$M University, College Station, TX
77843, USA }

\author{Tianjun Li}

\affiliation{George P. and Cynthia W. Mitchell Institute for
Fundamental Physics, Texas A$\&$M University, College Station, TX
77843, USA }

\affiliation{ Institute of Theoretical Physics, Chinese Academy of
Sciences, Beijing 100080, China}

\author{V. E. Mayes}

\affiliation{George P. and Cynthia W. Mitchell Institute for
Fundamental Physics, Texas A$\&$M University, College Station, TX
77843, USA }

\author{Dimitri V. Nanopoulos}

\affiliation{George P. and Cynthia W. Mitchell Institute for
Fundamental Physics, Texas A$\&$M University, College Station, TX
77843, USA }

\affiliation{Astroparticle Physics Group, Houston Advanced
Research Center (HARC), Mitchell Campus, Woodlands, TX 77381, USA}

\affiliation{Academy of Athens, Division of Natural Sciences, 28
Panepistimiou Avenue, Athens 10679, Greece }



\begin{abstract}

We briefly describe a three-family intersecting D6-brane model in
Type IIA theory on the $\mathbf{T^6/(\Z_2\times \Z_2)}$
orientifold with a realistic phenomenology. In this model, the
gauge symmetry can be broken down to the Standard Model (SM)
gauge symmetry close to the string scale, and the gauge coupling 
unification can be achieved. We calculate the supersymmetry breaking 
soft terms, and the corresponding low energy supersymmetric particle 
spectrum, which may be tested at the Large Hadron Collider (LHC). The observed
dark matter density may also be generated. Finally, we can explain
the SM quark masses and CKM mixings, and the tau lepton mass. The
neutrino masses and mixings may be generated via the seesaw
mechanism as well.

\end{abstract}


\preprint{ACT-01-07, MIFP-07-10}

\maketitle


{\bf Introduction~--}~During the last few years, intersecting
D-brane models on Type II orientifolds~\cite{JPEW}, where the
chiral fermions arise from the intersections of D-branes in the
internal space~\cite{bdl} and the T-dual description in terms of
magnetized D-branes~\cite{bachas} have been particularly
interesting~\cite{Blumenhagen:2005mu}. On Type IIA orientifolds
with intersecting D6-branes, a large number of non-supersymmetric
three-family Standard-like models and Grand Unified Theories
(GUTs) were constructed in early stages~\cite{Blumenhagen:2000wh}.
However, there generically existed uncancelled
Neveu-Schwarz-Neveu-Schwarz tadpoles in these models as well as
the gauge hierarchy problem. To solve these problems,
semi-realistic supersymmetric Standard-like and GUT models have
been constructed in Type IIA theory on the
$\mathbf{T^6/(\Z_2\times \Z_2)}$ orientifold~\cite{CSU,CLL} and
other backgrounds~\cite{ListSUSYOthers}. To stabilize the moduli
via supergravity fluxes, flux models on Type II orientifolds have
also been constructed~\cite{ListFlux,Chen:2006gd}. There are two
main constraints on supersymmetric D-brane model building: (1)
Ramond-Ramond (RR) tadpole cancellation conditions and (2) the
requirement for four-dimensional $N=1$ supersymmetric D-brane
configurations.

However, there are two serious problems in almost all
supersymmetric D-brane models: the absence of gauge coupling
unification at the string scale, and the rank one problem in the
Standard Model (SM) fermion Yukawa matrices. Thus, a comprehensive
phenomenological study of a concrete model from the string scale
to the electroweak scale has yet to be made. Although these problems can
be solved in the flux models of Ref.~\cite{Chen:2006gd} where the
RR tadpole cancellation conditions are relaxed, those models are in the
AdS vacua and the resulting flux induced superpotential for moduli
is too complicated.  Interestingly, we find that there is  one and
only one intersecting D6-brane model on the Type IIA
$\mathbf{T^6/(\Z_2\times \Z_2)}$ orientifold where the above
problems can be solved~\cite{CLL,Chen:2006gd}. Therefore, it is
desirable to study the phenomenological consequences of this model
in great detail.

{\bf Model Building~--}~We consider Type IIA string theory
compactified on a $\mathbf{T^6/(\Z_2\times \Z_2)}$
orientifold~\cite{CSU}. The $\mathbf{T^{6}}$ is a six-torus
factorized as $\mathbf{T^{6}} = \mathbf{T^2} \times \mathbf{T^2}
\times \mathbf{T^2}$ whose complex coordinates are $z_i$, $i=1,\;
2,\; 3$ for the $i^{th}$ two torus, respectively. The $\theta$ and
$\omega$ generators for the orbifold group $\mathbf{\Z_{2} \times \Z_{2}}$,
act on the complex coordinates of $\mathbf{T^6}$ as
\begin{eqnarray}
& \theta: & (z_1,z_2,z_3) \to (-z_1,-z_2,z_3)~,~ \nonumber \\
& \omega: & (z_1,z_2,z_3) \to (z_1,-z_2,-z_3)~.~\,
\label{orbifold}
\end{eqnarray}
The orientifold projection is implemented by gauging the symmetry
$\Omega R$, where $\Omega$ is world-sheet parity, and $R$ is given
by
\begin{eqnarray}
R: (z_1,z_2,z_3) \to ({\overline z}_1,{\overline z}_2,{\overline
z}_3)~.~\,    \label{orientifold}
\end{eqnarray}
Thus, there are four kinds of orientifold 6-planes (O6-planes) for
the actions $\Omega R$, $\Omega R\theta$, $\Omega R \omega$, and
$\Omega R\theta\omega$, respectively. There are two kinds of
complex structures consistent with orientifold projection for a
two torus: rectangular and tilted~\cite{CSU}. If we denote the
homology classes of the three cycles wrapped by the D6-brane
stacks as $n_P^i[a_i]+m_P^i[b_i]$ and $n_P^i[a'_i]+m_P^i[b_i]$
with $[a_i']=[a_i]+\frac{1}{2}[b_i]$ for the rectangular and
tilted tori respectively, we can label a generic one cycle by
$(n_P^i,l_P^i)$ in either case, where in terms of the wrapping
numbers $l_{P}^{i}\equiv m_{P}^{i}$ for a rectangular two torus
and $l_{P}^{i}\equiv 2\tilde{m}_{P}^{i}=2m_{P}^{i}+n_{P}^{i}$ for
a tilted two torus. Moreover, for a stack of $N$ D6-branes that
does not lie on one of the O6-planes, we obtain a $U(N/2)$ gauge
symmetry with three adjoint chiral superfields due to the orbifold
projections, while for a stack of $N$ D6-branes which lies on an
O6-plane, we obtain a $USp(N)$ gauge symmetry with three
anti-symmetric chiral superfields. Bifundamental chiral
superfields arise from the intersections of two different stacks
$P$ and $Q$  of D6-branes or from one stack $P$ and its $\Omega R$
image $P'$~\cite{CSU}.

We present the D6-brane configurations and intersection numbers of
the model in Table~\ref{MI-Numbers}, and the resulting spectrum in
Table~\ref{Spectrum}~\cite{CLL,Chen:2006gd}. We put the $a'$, $b$,
and $c$ stacks of D6-branes on the top of each other on the third
two torus, and as a resut there are additional vector-like
particles from $N=2$ subsectors.

\begin{table}[htb]
\footnotesize
\renewcommand{\arraystretch}{1.0}
\caption{D6-brane configurations and intersection numbers.}
\label{MI-Numbers}
\begin{center}
\begin{tabular}{|c||c|c||c|c|c|c|c|c|c|c|c|c|}
\hline
& \multicolumn{12}{c|}{$U(4)_C\times U(2)_L\times U(2)_R\times USp(2)^4$}\\
\hline \hline  & $N$ & $(n^1,l^1)\times (n^2,l^2)\times

(n^3,l^3)$ & $n_{S}$& $n_{A}$ & $b$ & $b'$ & $c$ & $c'$& 1 & 2 & 3 & 4 \\

\hline

    $a$&  8& $(0,-1)\times (1,1)\times (1,1)$ & 0 & 0  & 3 & 0 & -3 & 0 & 1 & -1 & 0 & 0\\

    $b$&  4& $(3,1)\times (1,0)\times (1,-1)$ & 2 & -2  & - & - & 0 & 0 & 0 & 1 & 0 & -3 \\

    $c$&  4& $(3,-1)\times (0,1)\times (1,-1)$ & -2 & 2  & - & - & - & - & -1 & 0 & 3 & 0\\

\hline

    1&   2& $(1,0)\times (1,0)\times (2,0)$ & \multicolumn{10}{c|}{$\chi_1=3,~
\chi_2=1,~\chi_3=2$}\\

    2&   2& $(1,0)\times (0,-1)\times (0,2)$ & \multicolumn{10}{c|}{$\beta^g_1=-3,~
\beta^g_2=-3$}\\

    3&   2& $(0,-1)\times (1,0)\times (0,2)$& \multicolumn{10}{c|}{$\beta^g_3=-3,~
\beta^g_4=-3$}\\

    4&   2& $(0,-1)\times (0,1)\times (2,0)$ & \multicolumn{10}{c|}{}\\

\hline

\end{tabular}

\end{center}

\end{table}

\begin{table}[htb]

\footnotesize

\renewcommand{\arraystretch}{1.0}

\caption{The chiral and vector-like superfields, and their quantum
numbers under the gauge symmetry $SU(4)_C\times SU(2)_L\times
SU(2)_R \times USp(2)_1 \times USp(2)_2 \times USp(2)_3 \times
USp(2)_4$.}

\label{Spectrum}

\begin{center}

\begin{tabular}{|c||c||c|c|c||c|c|c|}\hline

 & Quantum Number

& $Q_4$ & $Q_{2L}$ & $Q_{2R}$  & Field \\

\hline\hline

$ab$ & $3 \times (4,\overline{2},1,1,1,1,1)$ & 1 & -1 & 0  & $F_L(Q_L, L_L)$\\

$ac$ & $3\times (\overline{4},1,2,1,1,1,1)$ & -1 & 0 & $1$   & $F_R(Q_R, L_R)$\\

$a1$ & $1\times (4,1,1,2,1,1,1)$ & $1$ & 0 & 0  & \\

$a2$ & $1\times (\overline{4},1,1,1,2,1,1)$ & -1 & 0 & 0   & \\

$b2$ & $1\times(1,2,1,1,2,1,1)$ & 0 & 1 & 0    & \\

$b4$ & $3\times(1,\overline{2},1,1,1,1,2)$ & 0 & -1 & 0    & \\

$c1$ & $1\times(1,1,\overline{2},2,1,1,1)$ & 0 & 0 & -1    & \\

$c3$ & $3\times(1,1,2,1,1,2,1)$ & 0 & 0 & 1   &  \\

$b_{S}$ & $2\times(1,3,1,1,1,1,1)$ & 0 & 2 & 0   &  $T_L^i$ \\

$b_{A}$ & $2\times(1,\overline{1},1,1,1,1,1)$ & 0 & -2 & 0   & $S_L^i$ \\

$c_{S}$ & $2\times(1,1,\overline{3},1,1,1,1)$ & 0 & 0 & -2   & $T_R^i$  \\

$c_{A}$ & $2\times(1,1,1,1,1,1,1)$ & 0 & 0 & 2   & $S_R^i$ \\

\hline\hline

$ab'$ & $3 \times (4,2,1,1,1,1,1)$ & 1 & 1 & 0  & \\

& $3 \times (\overline{4},\overline{2},1,1,1,1,1)$ & -1 & -1 & 0  & \\

\hline

$ac'$ & $3 \times (4,1,2,1,1,1,1)$ & 1 &  & 1  & $\Phi_i$ \\

& $3 \times (\overline{4}, 1, \overline{2},1,1,1,1)$ & -1 & 0 & -1  &
$\overline{\Phi}_i$\\

\hline

$bc$ & $6 \times (1,2,\overline{2},1,1,1,1)$ & 0 & 1 & -1   & $H_u^i$, $H_d^i$\\

& $6 \times (1,\overline{2},2,1,1,1,1)$ & 0 & -1 & 1   & \\

\hline

\end{tabular}

\end{center}

\end{table}

The anomalies from three global $U(1)$s of $U(4)_C$, $U(2)_L$ and
$U(2)_R$ are cancelled by the Green-Schwarz mechanism, and the
gauge fields of these $U(1)$s obtain masses via the linear
$B\wedge F$ couplings. Thus, the effective gauge symmetry is
$SU(4)_C\times SU(2)_L\times SU(2)_R$. In order to break the gauge
symmetry, on the first torus, we split the $a$ stack of D6-branes
into $a_1$ and $a_2$ stacks with 6 and 2 D6-branes, respectively,
and split the $c$ stack of D6-branes into $c_1$ and $c_2$ stacks
with two D6-branes for each one. In this way, the gauge symmetry
is further broken to $ SU(3)_C\times SU(2)_L\times
U(1)_{I_{3R}}\times U(1)_{B-L}$. Moreover, the
$U(1)_{I_{3R}}\times U(1)_{B-L}$ gauge symmetry may be broken to
$U(1)_Y$ by giving vacuum expectation values (VEVs) to the
vector-like particles with the quantum numbers $({\bf { 1}, 1,
1/2, -1})$ and $({\bf { 1}, 1, -1/2, 1})$ under the $SU(3)_C\times
SU(2)_L\times U(1)_{I_{3R}} \times U(1)_{B-L} $ gauge symmetry
from $a_2 c_1'$ intersections~\cite{CLL,Chen:2006gd}.

Since the gauge couplings in the Minimal Supersymmetric Standard
Model (MSSM) are unified at the GUT scale $\sim2.4\times 10^{16}$
GeV, the additional exotic particles present in the model must
necessarily become superheavy. To accomplish this it is first
assumed that the $USp(2)_1$ and $USp(2)_2$ stacks of D6-branes lie
on the top of each other on the first torus, so we have two pairs
of  vector-like particles with $USp(2)_1\times USp(2)_2$ quantum
numbers $(2,2)$. These particles can break  $USp(2)_1\times
USp(2)_2$ down to the diagonal $USp(2)_{D12}$ near the string
scale, and then states arising from intersections $a1$ and $a2$
may obtain vector-like masses close to the string scale. Moreover,
we assume that the $T_R^i$ and $S_R^i$ obtain VEVs near the string
scale, and their VEVs satisfy the D-flatness of $U(1)_R$. To
preserve the D-flatness of $U(1)_L$, we  assume that the VEVs of
$S_L^i$ is TeV scale. We also assume that there exist various
suitable high-dimensional operators in the effective theory. With
$T_R^i$ and $S_R^i$, we can give the GUT-scale masses to the
particles from the intersections $c1$, $c3$, and $c_S$ via
high-dimensional operators. The remaining states and adjoint
chiral superfields may also obtain GUT-scale masses via
high-dimensional operators by the Higgs mechanism and from strong
dynamics since all of the $USp(2)_i$ have negative beta functions
as shown in Table \ref{MI-Numbers}~\cite{CLMN-L}. To have one pair
of light Higgs doublets, it is necessary to fine-tune the mixing
parameters of the Higgs doublets. In particular, the $\mu$ term
and the right-handed neutrino masses may be generated via the
following high-dimensional operators
\begin{eqnarray}
W \supset &&{{y^{ijkl}_{\mu}} \over {M_{\rm St}}} S_L^i S_R^j
H_u^k H_d^l + {{y^{mnkl}_{Nij}}\over {M^3_{\rm St}}} T_R^{m}
T_R^{n} \Phi_i \Phi_j  F_R^k  F_R^l ~,~\,
\end{eqnarray}
where $y^{ijkl}_{\mu}$ and $y^{mnkl}_{Nij}$ are Yukawa couplings,
and $M_{\rm St}$ is the string scale. Thus, the $\mu$ term is TeV
scale and the right-handed neutrino masses can be in the range
$10^{10-14}$ GeV for $y^{ijkl}_{\mu} \sim 1$ and $y^{mnkl}_{Nij}
\sim 10^{(-7)-(-3)}$.

{\bf Phenomenological Consequences~--}~In the string theory basis,
we have the dilaton $S$, three K\"ahler moduli $T^i$, and three
complex structure moduli $U^i$~\cite{Lust:2004cx}. The $U^i$ for
the present model are
\begin{equation}
U^1 = 3i~,~~ U^2 = i~,~~ U^3 = -1 + i~.~\,
\end{equation}

The corresponding moduli $s$, $t^i$ and $u^i$ in the supergravity
theory basis are related to the $S$, $T^i$ and $U^i$ moduli
by~\cite{Lust:2004cx}
\begin{eqnarray}
\mathrm{Re}\,(s)& =&
\frac{e^{-{\phi}_4}}{2\pi}\,\left(\frac{\sqrt{U^{1}_2\,
U^{2}_2\,U^3_2}}{|U^1U^2U^3|}\right)~,~
\mathrm{Re}(t^j)~=~\frac{i\alpha'}{T^j}~,~
\nonumber \\
\mathrm{Re}\,(u^j)& =&
\frac{e^{-{\phi}_4}}{2\pi}\left(\sqrt{\frac{U^{j}_2}
{U^{k}_2\,U^l_2}}\right)\; \left|\frac{U^k\,U^l}{U^j}\right| ~,~\,
\label{idb:eq:moduli}
\end{eqnarray}
where $\phi_4$ is the four-dimensional dilaton, $U^i_2$ is the
imaginary part of $U^i$, and $j\not=k\not=l\not=j$.

The holomorphic gauge kinetic function for a generic $P$ stack of
D6-branes which does not lie on one of O6-planes, is given
by~\cite{Lust:2004cx}
\begin{eqnarray}
f_P &=& {1\over 8}\left( 2 n_P^1\,n_P^2\,n_P^3\,s-
n_P^1\,l_P^2\,l_P^3\,u^1 \right.\nonumber\\&& \left.
-n_P^2\,l_P^1\,l_P^3\,u^2 -2n_P^3\,l_P^1\,l_P^2\,u^3\right)~.~\,
\end{eqnarray}
Thus, the gauge couplings for $SU(4)_C$, $SU(2)_L$ and $SU(2)_R$ 
in our model are unified at the string scale. For
simplicity, we negelect the little hierarchy between the string
scale and the GUT scale, which may be explained via threshold
corrections. Assuming the value of the unified gauge coupling in
the MSSM, we obtain
\begin{equation}
e^{-\phi_4} = 20.1~.~\,
\end{equation}
Thus, the string scale is $\sim2.1\times 10^{17}~{\rm GeV}$ for
$M_{\rm St}=\pi^{1/2} e^{\phi_4} M_{\rm Pl}$ where $ M_{\rm Pl}$
is the reduced Planck scale.

The K\"ahler metric for the chiral superfields from the
intersections of the $P$ and $Q$ stacks of D6-branes
is~\cite{Lust:2004cx}
\begin{equation}
\tilde{K} \supset e^{\phi_4+\gamma_E \sum_{i = 1}^3
\theta^{i}_{PQ}} \prod_{j=1}^3
\left[\sqrt{\frac{\Gamma(1-\theta^{i}_{PQ})}
{\Gamma(\theta^{i}_{PQ})}} (t^j +
\bar{t}^j)^{-\theta^{i}_{PQ}}\right], \nonumber
\end{equation}
where $\gamma_E$ is the Euler-Mascheroni constant, and
$\theta^{i}_{PQ}$ is the suitable positive angle between the $P$
and $Q$ stacks of D6-branes on the $i^{th}$ two torus in units of
$\pi$~\cite{CLMN-L}, and can be written as a function of $s$,
$u^i$, and the wrapping numbers for the $P$ and $Q$ stacks of
D6-branes.

The K\"ahler metric for the vector-like chiral superfields from
the intersections of the $P$ and $Q$ stacks of D6-branes that are
parallel on the $j^{th}$ two torus and intersect on the $k^{th}$
and $l^{th}$ two tori is given by~\cite{Lust:2004cx}
\begin{equation}
\tilde{K} \supset \left[ (s+\overline{s}) (u^j+\overline{u}^j)
(t^k+\overline{t}^k) (t^l+\overline{t}^l) \right]^{-1/2}~.~
\end{equation}

\begin{table}[htb]
\footnotesize

\renewcommand{\arraystretch}{1.0}

\caption{Supersymmetry breaking soft terms (in GeV) at the string scale.}

\label{Soft-Mass}

\begin{center}

\begin{tabular}{|c|c|c|c|c|c|c|c|}\hline

 $M_1$ & $M_2$ & $M_3$ & $m_{FL}$ &
$m_{FR}$ & $m_H$ & $A_Y$ \\ \hline

   477.4   & 279.1   & 987.8  & 1047   & 524.7   & 451.7
& 732.6     \\

\hline

\end{tabular}

\end{center}

\end{table}

For simplicity,  we assume that only the F terms of the complex
structure moduli $u^i$ break  supersymmetry and are parametrized
as follows
\begin{equation}
F^{u^i} = \sqrt{3}m_{3/2}(u^i + \bar{u}^i)\Theta_i~,~~{\rm
for}~i=1, 2, 3~,~\, \label{auxfields}
\end{equation}
where $m_{3/2}$ is the gravitino mass, and $\Theta_i$ are real
numbers and satisfy $\sum_{i=1}^3 |\Theta_i|^2 = 1$. Then, we can
calculate the gaugino masses ($M_i$), the universal scalar masses
$m_{FL}$ and $m_{FR}$ respectively for the left-handed and
right-handed SM fermions, the universal scalar mass $m_H$ for
Higgs fields $H^i_u$ and $H_d^i$, and the universal trilinear soft
term $A_Y$ at the string scale~\cite{Brignole:1997dp}. Choosing
$m_{3/2}=1100$ GeV, $\Theta_1=-0.6$, $\Theta_2=0.293$,
$\Theta_3=0.744$, ${\rm Re}t_1=1/6.6$, and ${\rm Re}t_2={\rm
Re}t_3=0.5$, we obtain the string-scale
supersymmetry breaking soft terms given
in Table~\ref{Soft-Mass}. Using the code {\tt
SuSpect}~\cite{Djouadi:2002ze}, we calculate the low energy
supersymmetric particle spectrum.  An example for $\tan\beta=46$
and positive $\mu$ is shown in Table~\ref{SSpectrum}. This
spectrum is consistent with all the known experiments and can be
tested at the LHC. Finally, using the code
{\tt MicrOMEGAs}~\cite{Belanger:2004yn}, we obtain a dark matter
density $\Omega h^2 = 0.117$ which is very close to the observed
value.

\begin{table}[htb]

\footnotesize

\renewcommand{\arraystretch}{1.0}

\caption{Low energy supersymmetric particles and their masses (in GeV).}

\label{SSpectrum}

\begin{center}

\begin{tabular}{|c|c|c|c|c|c|c|c|c|c|}\hline

$h^0$ & $H^0$ & $A^0$ & $H^{\pm}$ & ${\tilde g}$ & $\chi_1^{\pm}$
& $\chi_2^{\pm}$ & $\chi_1^{0}$ & $\chi_2^{0}$  \\ \hline

121.3  & 1016  & 1017   & 1020   & 2192 & 219.3   & 1406  & 199.3
&  219.4     \\ \hline

$\chi_3^{0}$ & $\chi_4^{0}$ & ${\tilde t}_1$ & ${\tilde t}_2$ &
${\tilde u}_1/{\tilde c}_1$ & ${\tilde u}_2/{\tilde c}_2$ &
${\tilde b}_1$ & ${\tilde b}_2$ & \\ \hline

1404 & 1405 & 1542 & 1912 & 1948 & 2144 & 1763 & 1915 & \\ \hline

${\tilde d}_1/{\tilde s}_1$ & ${\tilde d}_2/{\tilde s}_2$ &
${\tilde \tau}_1$ & ${\tilde \tau}_2$ & ${\tilde \nu}_{\tau}$ &
${\tilde e}_1/{\tilde \mu}_1$ & ${\tilde e}_2/{\tilde \mu}_2$ &
${\tilde \nu}_e/{\tilde \nu}_{\mu}$ & \\ \hline

1947 & 2146 & 234.4 & 1010 & 1000 & 550.2 & 1059 & 1056 & \\

\hline

\end{tabular}

\end{center}

\end{table}

{\bf The SM Fermion Masses and Mixings~--}~Because all the SM
fermions and Higgs fields arise from the intersections on the
first torus, we will only consider it for simplicity. The up-type
quark mass matrix $M^U$ at the GUT scale is~\cite{Cremades:2003qj}
\begin{equation}
c_0^U \left(\begin{array}{ccc}

A^U v_u^1 + E^U v_u^4 & B^U v_u^3 + F^U v_u^6 & D^U v_u^2 + C^U v_u^5 \\

C^U v_u^3 + D^U v_u^6 & A^U v_u^5 + E^U v_u^2 & B^U v_u^1 + F^U v_u^4 \\

F^U v_u^2 + B^U v_u^5 & C^U v_u^1 + D^U v_u^4 & A^U v_u^3 + E^U v_u^6

\end{array} \right), \nonumber \label{Yukawa generalU}
\end{equation}
where $v_u^i=\langle H^i_u \rangle$, and $c_0^U$ is a constant
which includes the quantum corrections and the contributions to
the Yukawa couplings from the second and third two tori. The theta
functions $A^U$, $B^U$, $C^U$, $D^U$, $E^U$, and $F^U$ are
\begin{eqnarray}
&& A^U \equiv \vartheta \left[\begin{array}{c} \epsilon^{U1}\\
\phi^{(1)} \end{array} \right] (\kappa^{(1)}),~ B^U \equiv
\vartheta \left[\begin{array}{c} \epsilon^{U1}+\frac{1}{3}\\
\phi^{(1)} \end{array} \right]
(\kappa^{(1)})~,  \nonumber \\
&&  C^U \equiv  \vartheta \left[\begin{array}{c}
\epsilon^{U1}-\frac{1}{3}\\  \phi^{(1)} \end{array} \right]
(\kappa^{(1)}),~ D^U \equiv \vartheta \left[\begin{array}{c}
\epsilon^{U1}+\frac{1}{6}\\  \phi^{(1)} \end{array} \right]
(\kappa^{(1)}), \nonumber   \\
&& E^U \equiv \vartheta \left[\begin{array}{c}
\epsilon^{U1}+\frac{1}{2}\\  \phi^{(1)} \end{array} \right]
(\kappa^{(1)}),~ F^U \equiv \vartheta \left[\begin{array}{c}
\epsilon^{U1}-\frac{1}{6}\\  \phi^{(1)} \end{array} \right]
(\kappa^{(1)}), \nonumber
\end{eqnarray}
where
\begin{eqnarray}
&& \epsilon^{U1} \equiv \frac{\epsilon_c^{U1} - \epsilon_b^{U1}
- 2\epsilon_a^{U1}}{6},~ \kappa^{(1)} \equiv \frac{6J^{(1)}}{\alpha'},~ \nonumber   \\
&& \phi^{(1)} =\;\;\; \theta_c^{(1)} - \theta_b^{(1)} -
2\theta_a^{(1)},~
\end{eqnarray}
where $\epsilon_a^{U1}$, $\epsilon_b^{U1}$ and
$\epsilon_c^{U1}$ respectively are the shifts of $a$, $b$, and $c$
stacks of D6-branes, $J^{(1)}$ is the K\"ahler modulus, and
$\theta_a^{(1)}$, $\theta_b^{(1)}$ and $\theta_c^{(1)}$ are the
Wilson line phases for the $a$, $b$, and $c$ stacks on the first
two torus, respectively.

At the GUT scale, the down-type quark mass matrix $M^D$ is
obtained from the above up-type quark mass matrix $M^U$ by
changing the upper index $U$ and lower
 index $u$ to $D$ and $d$, respectively.
The lepton mass matrix $M^L$ is obtained from $M^D$ by changing 
the upper index $D$ to $L$.

To generate the suitable SM fermion masses and mixings at the GUT
scale, we choose $\epsilon^{U1}=\epsilon^{L1}=0$,
$\epsilon^{D1}=0.061$, and $\kappa^{(1)}=39.6i$. And for Higgs
VEVs, we choose $v_1^u= 0.000266$, $v_2^u= 0.236$, $v_3^u= 0.999$,
$v_4^u= 0.981$,  $v_5^u= 0.00481$, $v_6^u= 0.0345$, $v_1^d=
0.00224$, $v_2^d= 0$, $v_3^d= 1.58$, $v_4^d=0$, $v_5^d= 0.0445$,
and $v_6^d= 0.0001$. Then, with suitable $c_0^U$, $c_0^D$, and
$c_0^L$, we obtain the SM fermion mass matrices at the GUT scale
\begin{eqnarray}
M^U \simeq m_t \left(\begin{array}{ccc}
0.000266 & 0.00109 & 0.00747  \\
0.00109 & 0.00481 & 0.0310 \\
0.00747 & 0.0310 & 0.999
\end{array} \right)~,~ \nonumber \\
M^D \simeq m_b \left(\begin{array}{ccc}
0.00141 & 0.000025 &  4 \times 10^{-6} \\
0.000155 & 0.028 & 0.0 \\
0.0 & 2.2 \times 10^{-7} & 1
\end{array} \right) ~,~ \nonumber \\
M^L \simeq m_{\tau} \left(\begin{array}{ccc}
0.00142 & 3.0 \times 10^{-6} &  2.8 \times 10^{-8} \\
3.0 \times 10^{-6} & 0.0282 & 1.4 \times 10^{-9} \\
2.8 \times 10^{-8} & 1.4 \times 10^{-9} & 1
\end{array} \right). \nonumber
\end{eqnarray}
The above mass matrices can produce the correct quark masses and
CKM mixings, and the correct $\tau$ lepton mass at the electroweak
scale~\cite{Fusaoka:1998vc}. The electron mass is about 6.5 times
larger that the expected value, while the muon mass is about
$40\%$ smaller. Similar to the GUTs~\cite{Nanopoulos:1982zm}, we
have roughly the wrong fermion mass relation $m_e/m_{\mu} \simeq
m_{d}/m_s$, and the correct electron and muon masses can be
generated via high-dimensional operators~\cite{CLMN-L}. Moreover,
the suitable neutrino masses and mixings can be generated via the
seesaw mechanism by choosing suitable Majorana mass matrix for the
right-handed neutrinos.

{\bf Conclusions~--}~We have briefly presented a three-family intersecting
D6-brane model where the gauge symmetry can be broken down to the
SM gauge symmetry and the gauge coupling unification can be
realized at the string scale. We have calculated the supersymmetry
breaking soft terms, and obtained the low energy supersymmetric
particle spectrum within the reach of the LHC. Our model may also
generate the observed dark matter density. Finally, we can explain
the SM quark masses and CKM mixings, and the tau lepton mass. The
neutrino masses and mixings may be generated via the seesaw
mechanism as well.

{\bf Acknowledgments~--}~This research was supported in part by
the Mitchell-Heep Chair in High Energy Physics (CMC), by the
Cambridge-Mitchell Collaboration in Theoretical Cosmology (TL),
and by the DOE grant DE-FG03-95-Er-40917 (DVN).


\end{document}